\documentclass[preprint,prd,eqsecnum,aps,epsf]{revtex4}

\begin{document}

\def\aprge{\buildrel > \over {_{\sim}}}
\def\aprle{\buildrel < \over {_{\sim}}}

\def\etal{{\it et.~al.}}
\def\ie{{\it i.e.}}
\def\eg{{\it e.g.}}

\def\bwt{\begin{widetext}}
\def\ewt{\end{widetext}}
\def\be{\begin{equation}}
\def\ee{\end{equation}}
\def\bea{\begin{eqnarray}}
\def\eea{\end{eqnarray}}
\def\bean{\begin{eqnarray*}}
\def\eean{\end{eqnarray*}}
\def\bary{\begin{array}}
\def\eary{\end{array}}
\def\bi{\bibitem}
\def\bit{\begin{itemize}}
\def\eit{\end{itemize}}

\def\lan{\langle}
\def\ran{\rangle}
\def\lra{\leftrightarrow}
\def\la{\leftarrow}
\def\ra{\rightarrow}
\def\dash{\mbox{-}}
\def\ol{\overline}

\def\ub{\ol{u}}
\def\db{\ol{d}}
\def\sb{\ol{s}}
\def\cb{\ol{c}}

\def\re{\rm Re}
\def\im{\rm Im}

\def \b{{\cal B}}
\def \ca{{\cal A}}
\def \ko{K^0}
\def \ok{\overline{K}^0}
\def \s{\sqrt{2}}
\def \st{\sqrt{3}}
\def \sx{\sqrt{6}}
%\begin{document}
%\begin{large}
\title{Hidden Local Symmetry and Chiral Effective Theory for Vector and Axial-vector Mesons}
\author{Yong-Liang Ma$^\dag$,  Qing Wang$^*$ and Yue-Liang Wu$^\dag$ }
\address{$^\dag$Institute of Theoretical Physics, Chinese Academy of Sciences,
 Beijing 100080, China\\
 $^*$Department of Physics, Tsinghua University, Beijing 100084, China }

\date{\today}
%%%%%%%%%%%%%%%%%%%%%%%%%%%%%%%
\begin{abstract}
In this paper, we present the full Lagrangian of mesons
(pseudoscalars, vectors and axial-vectors) to $O(p^4)$ by using
the explicit global chiral symmetry and hidden local symmetry in
the chiral limit. In this approach, we see that there are many
other terms besides the usual eleven terms given in the literature
from hidden local symmetry approach. Of particular, there are some
terms in our full results which are important for understanding
the vector meson dominance and $\pi-\pi$ scattering and providing
consistent predictions on the decay rates of
$a_1\rightarrow\gamma\pi$ and $a_1\rightarrow\rho\pi$ as well as
for constructing a consistent effective chiral Lagrangian with
chiral perturbation theory. It is likely that the structures of
the effective chiral Lagrangian for $O(p^4)$ given in the
literature by using hidden local symmetry are incomplete and
consequently the resulting couplings are not reliable. It is
examined that the more general effective chiral Lagrangian given
in present paper can provide more consistent predictions for the
low energy phenomenology of $\rho-a_1$ system and result in more
consistent descriptions on the low energy behavior of light flavor
mesons.
\end{abstract}
\maketitle

%%%%%%%%%%%%%%%%%%%%%%%%%%%%%%%%%%%%%%%%%%%%%%%%%%%%%%%%
\section{ INTRODUCTION}

 The strong interaction is believed to be described by $SU(3)$ gauge
theory. As an asymptotic free theory, it has successful
applications in high energy region (i.e., $E>1GeV$), but in low
energy region (i.e., $E<1GeV$), one cannot make ordinaty
perturbation calculations since, in this region, the coupling
constant becomes large. To describe the physics of strong
interaction in low energy region, one may develop some effective
theories which reflect the symmetries and symmetry breaking in
this energy region. In this note, we focus on the chiral effective
Lagrangian theory.

The basic idea of chiral effective Lagrangian theory can be
described as follows: Compared with the scale where the
nonperturbative effects become important, the masses of the
lightest three flavor quarks ($u, d$ and $s$) are smaller than the
QCD scale $\Lambda_{QCD}$. When neglecting the masses of these
quarks, QCD Lagrangian possesses an $U(3)_L\times U(3)_R$ flavor
chiral symmetry. The chiral effective theory was first proposed by
S.Weinberg in 1979 \cite{weinberg}, where the effective theory of
two light flavor quarks ($u$ and $d$) was built. Later on, the
effective theory of two ($u$ and $d$) and three ($u$, $d$ and $s$)
flavor cases was studied systematically up to $O(p^4)$ in
\cite{Gasser}. Besides the normal parity section, there are
anomalous sections in the effective theory
\cite{wess}\cite{witten}.

In addition to the pseudoscalar mesons, there are also vector and
axial-vector mesons in the meson spectrum. How to build an
effective theory of vector and axial-vector mesons was discussed
by many authors. In the literatures, many methods were used, such
as matter field method \cite{Ecker}, massive Yang-Mills method
\cite{massive1}\cite{massive2}, anti-symmetric tensor field method
\cite{Gasser}\cite{tensor2}, hidden local symmetry method
\cite{hidden1}\cite{M.Bando}\cite{M.Harada} and QCD Green function
approach\cite{Tsinghua}. In general, one should also consider the
light scalar mesons which have been shown\cite{DWu} to play an
important role for understanding the dynamically spontaneous
symmetry breaking of the chiral symmetry $U(3)_L\times U(3)_R$. Of
particular, a chiral effective Lagrangian with scalars can be
derived from integrating out the quark and gluon fields by using a
new symmetry-preserving loop regularization method\cite{ylw}, and
the gap equations have been found to be resulted from minimal
conditions of effective potential for the scalar fields. It then
predicts the existence of $\sigma$ and $\kappa$ scalars as the
nonet scalar mesons which can be regarded as composite Higgs
bosons with a consistent mass spectra to the current experimental
data\cite{DWu}. For simplicity, we will not include the scalar
mesons in this paper.

The hidden local symmetry method is based on a popular idea that
the nonlinear $\sigma$ model based on the manifold $G/H$ is gauge
equivalent to the $\sigma$ model based on $G\times H_{local}$ and
the gauge bosons correspond to the local symmetry can be regarded
as composite bosons. In our present consideration, we will use the
extended hidden local symmetry where $G=U(3)_L\times U(3)_R$ for
three light flavor ($u, d$ and $s$) case. This model is gauge
equivalent to the nonlinear $\sigma$ model based on the manifold
$G/H$ \cite{M.Bando}\cite{M.Bando1}. Of course, there are maybe
contributions to the coefficients of the nonlinear $\sigma$ model
from the Yang-Mills-type self-interaction of the hidden
symmetry\cite{duality}, but we will not consider this case in this
note. In the hidden local symmetry method, the vector and
axial-vector mesons are treated as combinations of the dynamical
gauge bosons of hidden local symmetry
$G_{local}=\hat{U}(3)_L\times\hat{U}(3)_R$ as suggested in
\cite{M.Bando}. But after a careful check, we will see that there
are many terms including three important terms to $O(p^4)$ were
missed in the literature \cite{M.Bando}, it is these three
important terms that can cancel the strong momentum dependence of
the $\rho-\pi-\pi$ coupling $f_{\rho\pi\pi}$ and also it is these
three terms that can ensure the $\rho-$ meson dominance in
$a_1\rightarrow\gamma\pi$ decay and result in consistent
predictions on the decay rates $a_1\rightarrow\gamma\pi$ and
$a_1\rightarrow\rho\pi$. Of particular, these new terms play an
important role for understanding the $\pi-\pi$ scattering
\cite{Gasser}\cite{pion-pion}\cite{libian}, or more generally, the
meson-meson scattering.

The paper is organized as follows, in section II, we will give a
simple but complete description of hidden local symmetry. In
section III, after list the fourteen important terms of the
effective chiral Lagrangian, we choose a special gauge, i.e.,
unitary gauge, and explicitly present a gauged Lagrangian
explicitly. In section IV, it is shown that with an appropriate
gauge fixing condition for the hidden local symmetry, fourteen
parameters appearing in the more general effective Lagrangian
based on the explicit global chiral symmetry and hidden local
chiral symmetry can be uniquely extracted when comparing it with
the effective Lagrangian of chiral perturbation theory. The
relevant low energy phenomenologes of $\rho-a_1$ system, such as
universality of the $\rho-$meson coupling, vector meson dominance,
the $\rho-\pi-\pi$ coupling $f_{\rho\pi\pi}$, the KSFR relation
$m^2_\rho=f^2_{\rho\pi\pi}f^2_\pi/2$, etc., are discussed in
section V. Our conclusions and remarks are presented in section
VI. The full Lagrangian up to $O(p^4)$ is presented in the
Appendix.
%%%%%%%%%%%%%%%%%%%%%%%%%%%%%%%%%%%%%%%%%%%%%%%%%%
\section{HIDDEN LOCAL SYMMETRY}

In chiral limit, the vector and axial-vector mesons can not be
introduced as gauge bosons via gauging the above global chiral
symmetry $G_{global}=U(3)_L\times U(3)_R$, otherwise there exists,
according to the Higgs mechanism, no independent degrees of
freedoms for the Goldstone-like pseudoscalar mesons. On the other
hand, the chiral gauge boson couplings to the light quarks must be
invariant under the transformation of the global chiral symmetry
$G_{global}$ as the original QCD theory does. It is then motivated
to introduce hidden local chiral symmetry
$G_{local}=\hat{U}(3)_L\times \hat{U}(3)_R$ associated with the
chiral gauge bosons $\hat{A}_L$ and $\hat{A}_R$. After the
spontaneous breaking of the global chiral symmetry $G_{global}$,
the Goldstone-like pseudoscalar mesons are generated, the chiral
gauge bosons associated with the hidden local gauge symmetry also
turn out to be vector and axial-vector mesons via an appropriate
choice of the gauge transformation of the hidden local symmetry
$G_{local}$. Such a gauge choice breaks the hidden local chiral
symmetry and generates the masses of the vector and axial-vector
mesons. In this paper, we are limited to consider the case of
chiral limit and will not discuss the gauge anomalous section.

Let's begin with introducing the necessary fields for constructing
the chiral Lagrangian which are covariant under the global chiral
symmetry $G_{global}$. The chiral Lagrangian is expected to
describe the Goldstone-like pseudoscalar mesons, vector mesons and
axial-vector mesons which arise from the gauge bosons of the local
chiral symmetry $G_{local}$.

Corresponding to the global chiral symmetry $G_{global}$, we
introduce the local chiral symmetry $G_{local}$. In this case, we
have the nonlinear chiral fields $\hat{\xi}_L(x)\in U(3)_L\times
\hat{U}(3)_L$ and $\hat{\xi}_R(x)\in U(3)_R\times \hat{U}(3)_R$,
which transform as
\begin{eqnarray}
&&\hat{\xi}_L(x)\rightarrow
g_L\hat{\xi}_L(x)G_L^\dag(x);~~~~~~g_L\in
U(3)_L,~~~~G_L\in\hat{U}(3)_L\\
&&\hat{\xi}_R(x)\rightarrow
g_R\hat{\xi}_L(x)G_R^\dag(x);~~~~~~g_R\in
U(3)_R,~~~~G_R\in\hat{U}(3)_R
\end{eqnarray}

We also have nonlinear chiral field $\xi_M(x)\in G_{local}$, its
transformation property is
\begin{eqnarray}
\xi_M(x)\rightarrow
G_L(x)\xi_M(x)G_R^\dag(x);~~~~~~(G_L(x),G_R(x))\in
\hat{U}(3)_L\times \hat{U}(3)_R
\end{eqnarray}

With the above nonlinear chiral fields, we can construct nonlinear
field $U(x)\in G_{global}$ as follows
\begin{eqnarray}
U(x)\equiv \hat{\xi}_L(x)\xi_M(x)\hat{\xi}^\dag_R(x)\label{definU}
\end{eqnarray}
and its transformation property under the full group
$G_{global}\times G_{local}$ is
\begin{eqnarray}
U(x)\rightarrow g_L U(x)g_R^\dag;~~~~(g_L,g_R)\in G_{global}
\end{eqnarray}

From the above, we can also see that the transformation properties
of gauge fields $\hat{A}_L$ and $\hat{A}_R$ corresponding to local
chiral symmetry $G_{local}$ are
\begin{eqnarray}
&&\hat{A}_L\rightarrow\hat{A}_L^\prime=G_L(x)(\hat{A}_L+i\partial)G_L^\dag(x)\\
&&\hat{A}_R\rightarrow\hat{A}_R^\prime=G_R(x)(\hat{A}_R+i\partial)G_R^\dag(x)
\end{eqnarray}

Similarly, we can construct chiral gauge bosons
\begin{eqnarray}
&&a_L(x)=\hat{\xi}_L(x)(\hat{A}_L(x)+i\partial)\hat{\xi}_L^\dag(x)\equiv\hat{\xi}_L(x)iD\hat{\xi}_L^\dag(x)\\
&&a_R(x)=\hat{\xi}_R(x)(\hat{A}_R(x)+i\partial)\hat{\xi}_R^\dag(x)\equiv\hat{\xi}_R(x)iD\hat{\xi}_R^\dag(x)
\end{eqnarray}
their transformation properties under the full chiral symmetry
$G_{global}\times G_{local}$ are
\begin{eqnarray}
a_L(x)\rightarrow g_La_L(x)g_L^\dag,~~~~~~a_R(x)\rightarrow
g_Ra_R(x)g_R^\dag
\end{eqnarray}

The field strengths corresponding to local chiral symmetry are
\begin{eqnarray}
&&\hat{F}_L^{\mu\nu}=\partial^\mu \hat{A}_L^\nu-\partial^\nu
\hat{A}_L^\mu-i[\hat{A}^\mu_L,\hat{A}^\nu_L]\nonumber\\
&&\hat{F}_R^{\mu\nu}=\partial^\mu \hat{A}_R^\nu-\partial^\nu
\hat{A}_R^\mu-i[\hat{A}^\mu_R,\hat{A}^\nu_R]
\end{eqnarray}
then, the field strengths of chiral gauge bosons corresponding to
global chiral symmetry are
\begin{eqnarray}
&&F_L^{\mu\nu}=\partial^\mu a_L^\nu-\partial^\nu
a_L^\mu-i[a^\mu_L,a^\nu_L]=\hat{\xi}_L(x)\hat{F}_L^{\mu\nu}\hat{\xi}^\dag_L(x)\nonumber\\
&&F_R^{\mu\nu}=\partial^\mu a_R^\nu-\partial^\nu
a_R^\mu-i[a^\mu_R,a^\nu_R]=\hat{\xi}_R(x)\hat{F}_R^{\mu\nu}\hat{\xi}^\dag_R(x)
\end{eqnarray}
All the above field strengths are covariant
\begin{eqnarray}
&&F_L^{\mu\nu}\rightarrow
g_LF_L^{\mu\nu}g_L^\dag,~~~~~~F_R^{\mu\nu}\rightarrow
g_RF_R^{\mu\nu}g_R^\dag\nonumber\\
&&\hat{F}_L^{\mu\nu}\rightarrow
G_L\hat{F}_L^{\mu\nu}G_L^\dag,~~~~~~\hat{F}_R^{\mu\nu}\rightarrow
G_R\hat{F}_R^{\mu\nu}G_R^\dag
\end{eqnarray}
Similarly, we can also construct gauge fields as follows
\begin{eqnarray}
&&-\hat{a}_L(x)\equiv\xi_M(x)iD\xi_M^\dag(x)=\xi_M(x)(i\partial+\hat{A}_R(x))\xi^\dag_M(x)-\hat{A}_L(x)\\
&&-\hat{a}_R(x)\equiv\xi^\dag_M(x)iD\xi_M(x)=\xi^\dag_M(x)(i\partial+\hat{A}_L(x))\xi_M(x)-\hat{A}_R(x)=\xi_M^\dag(x)\hat{a}_L(x)\xi_M(x)
\label{definehata}
\end{eqnarray}
they are also covariant under local chiral symmetry
\begin{eqnarray}
\hat{a}_L(x)\rightarrow
G_L(x)\hat{a}_L(x)G_L^\dag(x),~~~~~~\hat{a}_R(x)\rightarrow
G_R(x)\hat{a}_R(x)G_R^\dag(x)
\end{eqnarray}

In above, we have defined a set of quantities, but in the sense of
gauge fields, there are only two kinds of independent quantities:
quantities (hatted quantities) transform according to local chiral
symmetry and quantities (unhatted quantities) transform according
to global chiral symmetry. They are equivalent in expressing gauge
fields since there are only two kinds of gauge fields in our
theory. The differences among them are chiral rotated angles. By
using these gauge fields and pseudoscalar fields, we can construct
chiral, $C, P$ and $T$ invariant lagrangian which consists of
pseudoscalar mensons, vector mesons and axial-vector mesons. The
Lagrangian, which will be constructed below, should be invariant
under the transformations of global chiral symmetry $U(3)_L\times
U(3)_R$ with the local chiral symmetry $\hat{U}(3)_L\times
\hat{U}(3)_R$ appearing as a hidden symmetry.

%%%%%%%%%%%%%%%%%%%%%%%%%%%%%%%%%%%%%%%%%%%%%%%%%%

\section{THE EFFECTIVE LAGRANGIAN OF
VECTOR, AXIAL-VECTOR AND PEUSODOSCALAR MESONS}

To construct the Lagrangian, we should take independent quantities
from those defined above. By analyzing their transformation
properties, the independent quantities may be chosen as follows
\begin{eqnarray}
a_{L\mu},~~~~Ua_{R\mu}U^\dag,~~~~\hat{\xi}_L\hat{a}_{L\mu}\hat{\xi}_L^\dag,
\end{eqnarray}
They are transforming as $A\rightarrow g_LAg_L^\dag$ with $A$
denotes the above three quantities.

Thus the $O(p^2)$ Lagrangian can be constructed as
\begin{eqnarray}
{\cal L}^2&=&{\cal L}^2_a+{\cal L}^2_b+{\cal L}^2_c+{\cal
L}^2_d\nonumber\\
{\cal L}^2_a&=&aTr[a_{L\mu}+Ua_{R\mu}U^\dag]^2\nonumber\\
{\cal L}^2_b&=&bTr[a_{L\mu}-Ua_{R\mu}U^\dag]^2\nonumber\\
{\cal L}^2_c&=&cTr[\hat{\xi}_L\hat{a}_{L\mu}\hat{a}_L^{\mu}\hat{\xi}_L^\dag]\nonumber\\
{\cal
L}^2_d&=&dTr[(a_{L\mu}-Ua_{R\mu}U^\dag)-\hat{\xi}_L\hat{a}_{L\mu}\hat{\xi}_L^\dag]^2
\end{eqnarray}
where $a, b, c$ and $d$ are constants and will be fixed later.

When rewriting the lagrangian in the explicit form of chiral angle
and covariant derivative and redefining the constants $a, b, c$
and $d$, we have
\begin{eqnarray}
{\cal L}^2_a&=&-a(f^2_\pi/16)Tr[\hat{\xi}_LD_\mu\hat{\xi}_L^\dag+(\hat{\xi}_L\xi_M)(D_\mu\hat{\xi}_R^\dag)\hat{\xi}_R(\hat{\xi}_L\xi_M)^\dag]^2\nonumber\\
{\cal L}^2_b&=&-b(f^2_\pi/16)Tr[\hat{\xi}_LD_\mu\hat{\xi}_L^\dag-(\hat{\xi}_L\xi_M)(D_\mu\hat{\xi}_R^\dag)\hat{\xi}_R(\hat{\xi}_L\xi_M)^\dag]^2\nonumber\\
{\cal L}^2_c&=&-c(f^2_\pi/16)Tr[\xi^\dag_MD_\mu\xi_M]^2\nonumber\\
{\cal
L}^2_d&=&-d(f^2_\pi/16)Tr[\hat{\xi}_LD_\mu\hat{\xi}_L^\dag+(\hat{\xi}_L\xi_M)(D_\mu\hat{\xi}_R^\dag)\hat{\xi}_R(\hat{\xi}_L\xi_M)^\dag-\hat{\xi}_L(\xi_MD_\mu\xi_M^\dag)\hat{\xi}_L^\dag]^2
\end{eqnarray}
which are $O(p^2)$ Lagrangian with $f_\pi$ the decay constant.

To construct $O(p^4)$ Lagrangian, let's define some quantities by
using the above independent quantities and confirm their parity
$(P)$ properties,

\begin{eqnarray}
&&P:~~~~a_{+\mu}\equiv(a_{L\mu}+Ua_{R\mu}U^\dag)\rightarrow U^\dag
a_{+\mu}U\label{aparallel}\\
&&P:~~~~a_{-\mu}\equiv(a_{L\mu}-Ua_{R\mu}U^\dag)\rightarrow-U^\dag
a_{-\mu}U\label{aperp}\\
&&P:~~~~\hat{a}_{-\mu}\equiv\hat{\xi}_L\hat{a}_{L\mu}\hat{\xi}_L^\dag\rightarrow-U^\dag
\hat{a}_{-\mu}U\label{hat}\\
&&P:~~~~V_{\mu\nu}\equiv
F_{\mu\nu}^L+UF_{\mu\nu}^RU^\dag\rightarrow U^\dag V_{\mu\nu}U\label{parityV}\\
&&P:~~~~A_{\mu\nu}\equiv
F_{\mu\nu}^L-UF_{\mu\nu}^RU^\dag\rightarrow -U^\dag
A_{\mu\nu}U\label{parityA}
\end{eqnarray}

From the above discussions, the $O(p^4)$ Lagrangian can easily be
constructed. A complete Lagrangian is presented in Appendix. Here
we focus ${\cal L}^2$ and the following ten relevant important
terms
\begin{eqnarray}
{\cal L}^4&=&{\cal L}^4_k+{\cal L}^4_{\hat{a}}+{\cal L}^4_F\nonumber\\
{\cal
L}^4_k&=&-{1\over4g^2_G}Tr(F_{L\mu\nu}F^{\mu\nu}_L+F_{R\mu\nu}F^{\mu\nu}_R)=-{1\over4g_G^2}Tr(\hat{F}_{L\mu\nu}\hat{F}^{\mu\nu}_L+\hat{F}_{R\mu\nu}\hat{F}^{\mu\nu}_R)\\\label{dynamicalterm}
{\cal L}^4_{\hat{a}}&=&\alpha(1/12g^2_G)Tr[\hat{\xi}_L(D_\mu\hat{a}_{L\nu})(D^\mu\hat{a}_L^\nu)\hat{\xi}_L^\dag]\nonumber\\
&&+\beta(1/12g^2_G)Tr[\hat{\xi}_L\hat{a}_{L\mu}\hat{a}_{L\nu}\hat{a}_L^\mu\hat{a}_L^\nu\hat{\xi}_L^\dag]\nonumber\\
&&+\gamma(1/12g^2_G)Tr[(\hat{\xi}_L\hat{a}_{L\mu}\hat{a}_L^\mu\hat{\xi}_L^\dag)^2]\nonumber\\
{\cal
L}^4_F&=&\alpha_1(-i/g_G^2)Tr[a_{L\mu}a_{L\nu}F^{L\mu\nu}+a_{R\mu}a_{R\nu}F^{R\mu\nu}]\nonumber\\
&&+\alpha_2(-i/g_G^2)Tr[Ua_{R\mu}a_{R\nu}U^\dag F^{L\mu\nu}+a_{L\mu}a_{L\nu}UF^{R\mu\nu}U^\dag]\nonumber\\
&&+\alpha_3(+i/2g_G^2)Tr[a_{L\mu}Ua_{R\nu}U^\dag F^{L\mu\nu}+a_{R\mu}U^\dag a_{L\nu}UF^{R\mu\nu}]+H.c.\nonumber\\
&&+\alpha_4(-i/4g_G^2)Tr[\hat{\xi}_L\hat{a}_{L\mu}\hat{a}_{L\nu}\hat{\xi}_L^\dag F^{L\mu\nu}+\hat{\xi}_R\xi_M^\dag\hat{a}_{L\mu}\hat{a}_{L\nu}\xi_M\hat{\xi}_R^\dag F^{R\mu\nu}]\nonumber\\
&&+\alpha_5(+i/4g_G^2)Tr[a_{L\mu}\hat{\xi}_L\hat{a}_{L\nu}\hat{\xi}_L^\dag F^{L\mu\nu}+a_{R\mu}\hat{\xi}_R\hat{a}_{R\nu}\hat{\xi}_R^\dag F^{R\mu\nu}]+H.c.\nonumber\\
&&+\alpha_6(-i/4g_G^2)Tr[Ua_{R\mu}U^\dag\hat{\xi}_L\hat{a}_{L\nu}\hat{\xi}_L^\dag F^{L\mu\nu}-U^\dag a_{L\mu}U\hat{\xi}_R\hat{a}_{R\nu}\hat{\xi}_R^\dag F^{R\mu\nu}]+H.c.\nonumber\\
\end{eqnarray}

For comparison, the coupling constants are taken in terms of the
same notations as the ones in ref.\cite{M.Bando}\cite{M.Bando1}
except three additional interaction terms of $O(p^4)$ which have
been missed in \cite{M.Bando}\cite{M.Bando1} and will be found to
be very important for understanding the $\rho\pi\pi$ coupling
$g_{\rho\pi\pi}$ and the decay rates of $a_1\rightarrow\rho\pi$
and $a_1\rightarrow\gamma\pi$. It is seen that there are fourteen
unknown coupling constants: $a, b, c, d, g_G, \alpha, \beta,
\gamma$ and $\alpha_i (i=1,\cdots,6)$. In general, they need to be
determined via experimental processes and the success of current
algebra can also fix some of the couplings. It was shown in
\cite{M.Bando}\cite{M.Bando1} that the following choice of the
parameters seem to be consistent with the low energy phenomenology
and current algebra

\begin{eqnarray}
&&a=b=c=2, d=0\\
&&\alpha_1=\alpha_2=\alpha_3=0,~~~~-\alpha_4=\alpha_5=\alpha_6=1\\
&&\alpha, \beta, \gamma - missed,~~~~or~~~~\alpha=\beta=\gamma=0
\end{eqnarray}

Note that the values of this set of parameters were
phenomenologically suggested including the three terms $\alpha,
\beta$ and $\gamma$. The three additional terms are introduced at
the first time in this paper from hidden local symmetry. We will
discuss their values in the next section in detail.

Since the physics is independent of Hidden Local symmetry, we can
choose any appropriate gauges for the local symmetry to obtain the
effective Lagrangian for describing the low energy dynamics of
QCD. For convenient, we choose the following gauge transformations
of $G_{L,R}(x)$ which are the same as \cite{M.Bando}, so that
\begin{eqnarray}
&&\xi_M(x)\rightarrow G_L(x)\xi_M(x)G_R^\dag=1\\
&&\hat{\xi}_L(x)\rightarrow\hat{\xi}_L(x)G_L^\dag(x)=\xi_L(x)=\xi(x)=e^{i\Pi(x)/f_\pi}\\
&&\hat{\xi}_R(x)\rightarrow\hat{\xi}_R(x)G_R^\dag(x)=\xi_R(x)=\xi^\dag(x)=e^{-i\Pi(x)/f_\pi}\\
&&U(x)=\xi_L(x)\xi_R^\dag(x)=\xi^2(x)=e^{i2\Pi(x)/f_\pi}
\end{eqnarray}
where $\Pi(x)=\Pi^a\lambda^a$ is the nonet Goldstone-like
pseudoscalar. In our convention, $f_\pi=186MeV$. With this gauge,
then we have
\begin{eqnarray}
\hat{a}_R(x)=-\hat{a}_L=\hat{A}_R-\hat{A}_L=\xi_L^\dag(-iDU)\xi_R=\xi_R^\dag(iDU^\dag)\xi_L
\end{eqnarray}
where
\begin{eqnarray}
DU=\partial U+iUa_R-ia_LU
\end{eqnarray}

It is seen that the above choice of gauge condition is a kind of
unitary gauge corresponding to the broken down of the hidden local
chiral symmetry.

It will also be useful to decompose the chiral gauge fields
$\hat{A}_L$ and $\hat{A}_R$ into two parts
\begin{eqnarray}
\hat{A}_L(x)=A_L+L_\xi(x),~~~~\hat{A}_R(x)=A_R+R_\xi(x)
\end{eqnarray}
where $A_L(x)$ and $A_R(x)$ are the covariant parts associated
with the gauge bosons $a_L(x)$ and $a_R(x)$, while $L_\xi(x)$ and
$R_\xi(x)$ are the pure gauge parts associated with the
Goldstone-like pseudoscalars contained in the nonlinear chiral
fields $\hat{\xi}_L(x)$ and $\hat{\xi}_R(x)$
\begin{eqnarray}
&&A_L(x)=\xi^\dag_L(x)a_L(x)\xi_L(x)\equiv V(x)-A(x)\\
&&L_\xi(x)=\xi^\dag_L(x)i\partial\xi_L(x)\equiv
V_\xi(x)-A_\xi(x)\\
&&A_R(x)=\xi^\dag_R(x)a_R(x)\xi_R(x)\equiv V(x)+A(x)\\
&&R_\xi(x)=\xi^\dag_R(x)i\partial\xi_R(x)\equiv V_\xi(x)+A_\xi(x)
\end{eqnarray}
we can explicitly get
\begin{eqnarray}
2A_\xi&=&\xi^\dag_L(-i\partial U)\xi_R
\end{eqnarray}

In above gauge, we get the effective Lagrangian which possesses
the global $U(3)_L\times U(3)_R$ symmetry.

The $O(p^2)$ Lagrangian becomes
\begin{eqnarray}
{\cal L}^2&=&(a+b)(f^2_\pi/16)Tr[a_{L\mu}^2+a_{R\mu}^2]+2(a-b)(f^2_\pi/16)Tr[a_{L\mu}Ua_{R\mu}U^\dag]\nonumber\\
&&+c(f^2_\pi/16)Tr[D_\mu UD^\mu U^\dag
]+d(f^2_\pi/16)Tr[\partial_\mu U\partial^\mu U^\dag]\label{p2}
\end{eqnarray}

The $O(p^4)$ Lagrangian becomes
\begin{eqnarray}
{\cal L}_4&=&{\cal L}^4_k+{\cal L}^4_{\hat{a}}+{\cal L}^4_F+\cdots\nonumber\\
 {\cal
L}^4_k&=&-{1\over4g^2_G}Tr(F_{L\mu\nu}F^{\mu\nu}_L+F_{R\mu\nu}F^{\mu\nu}_R)=-{1\over4g^2_G}Tr(\hat{F}_{L\mu\nu}\hat{F}^{\mu\nu}_L+\hat{F}_{R\mu\nu}\hat{F}^{\mu\nu}_R)\\\label{dynamicalterm}
{\cal L}^4_{\hat{a}}&=&\alpha(1/12g^2_G)Tr[D_\mu D_\nu UD^\mu D^\nu U^\dag]\nonumber\\
&&+\beta(1/12g^2_G)Tr[D_\mu UD_\nu U^\dag D^\mu U D^\nu U^\dag]\nonumber\\
&&+\gamma(1/12g^2_G)Tr[D_\mu UD^\mu U^\dag D_\nu UD^\nu U^\dag]\nonumber\\
{\cal
L}^4_F&=&\alpha_1(-i/g_G^2)Tr[a_{L\mu}a_{L\nu}F^{L\mu\nu}+a_{R\mu}a_{R\nu}F^{R\mu\nu}]\nonumber\\
&&+\alpha_2(-i/g_G^2)Tr[a_{R\mu}a_{R\nu}U^\dag F^{L\mu\nu}U+a_{L\mu}a_{L\nu}UF^{R\mu\nu}U^\dag]\nonumber\\
&&+\alpha_3(+i/2g_G^2)Tr[a_{L\mu}Ua_{R\nu}U^\dag F^{L\mu\nu}+a_{R\mu}U^\dag a_{L\nu}UF^{R\mu\nu}]+H.c.\nonumber\\
&&+\alpha_4(-i/4g_G^2)Tr[D_\mu UD_\nu U^\dag F^{L\mu\nu}+D_\mu U^\dag D_\nu U F^{R\mu\nu}]\nonumber\\
&&+\alpha_5(+i/4g_G^2)Tr[a_{L\mu}iD_\nu U U^\dag F^{L\mu\nu}-a_{R\mu} iD_\nu U^\dag UF^{R\mu\nu}]+H.c.\nonumber\\
&&+\alpha_6(-i/4g_G^2)Tr[Ua_{R\mu}iD_\nu U^\dag F^{L\mu\nu}-U^\dag
a_{L\mu}iD_\nu U F^{R\mu\nu}]+H.c.\label{p4}
\end{eqnarray}

We will see below that it is this form of effective chiral
Lagrangian that enables us to compare it with the one derived from
effective chiral theory and chiral perturbation theory. This is
because they possess the same global chiral symmetry $G_{global}$
in the chiral limit. It then allows us to fix the fourteen
parameters in terms of two parameters introduced in the effective
chiral theory of mesons in the large $N_c$ approach.

%%%%%%%%%%%%%%%%%%%%%%%%%%%%%%%%%%%%%%%%%%%%%%%%%%%%%%

\section{14-PARAMETERS IN CHIRAL LAGRANGIAN
OF HIDDEN LOCAL SYMMETRY}

So far, the effective chiral lagrangian based on the global chiral
symmetry and local hidden symmetry breaking has been presented.
Considering the appropriate gauge selection mentioned in last
section, we can fix the 14-parameters by comparing them with the
ones of chiral perturbation theory given in \cite{libian}. It is
easy to check that the parameters are fixed to be

\begin{eqnarray}
&&a=b=\frac{m_0^2}{f^2_\pi}=\frac{g^2m^2_\rho}{f^2_\pi},~~~~c=\frac{6g^2m^2}{f^2_\pi}=\frac{F^2}{f^2_\pi},~~~~d=0\nonumber\\
&&\alpha=2\beta=-\gamma=\frac{N_c}{2(\pi
g)^2},~~~~g^2_G=\frac{4}{g^2}\nonumber\\
&&\alpha_1=\alpha_2=\alpha_3=\alpha_5=\alpha_6=0,~~~~\alpha_4=\frac{N_c}{2(\pi
g)^2}=\alpha\label{chpara}
\end{eqnarray}
with the redefinition
\begin{eqnarray}
g^2={1\over6}\frac{F^2}{m^2}
\end{eqnarray}
where $m$ and $m_0$ are free parameters.

To define the physical meson states in the mass eigenstates, one
needs to normalize the kinetic terms and redefine the
pseudoscalars and axial-vectors due to the mixing term
$a^\mu(x)\partial_\mu \pi(x)$, which leads to
\begin{eqnarray}
&&f^2_\pi=F^2\bigg(1-\frac{2c}{a+b+2c}\bigg)=F^2\bigg(1-\frac{6m^2}{m^2_\rho+6m^2}\bigg)\nonumber\\
&&m^2_\rho=m^2_o/g^2\label{normal}
\end{eqnarray}

Comparing the above parameters with the one fixed via the low
energy phenomenology, we can get the following conclusions:

    (1).~~For the terms in the effective Lagrangian up to
    the $O(p^2)$, both effective chiral Lagrangian approach and hidden
    local symmetry approach provide a consistent determination for the
    four parameters. It is interesting to note that once the vector mass is dynamically generated and takes the value
    \begin{eqnarray}
    m^2_\rho=6m^2
    \end{eqnarray}
   we have from ($\ref{chpara}$) and ($\ref{normal}$)
    \begin{eqnarray}
    F^2=2f^2_\pi,~~~~~~~~a=b=c=2
    \end{eqnarray}
    which agree well with the conclusions obtained from the
    current algebra and phenomenology analysis in the hidden
    symmetry approach \cite{M.Bando}\cite{M.Bando1}.

    (2).~~From the terms in $O(p^4)$, we noticed that: (i)
    there are ten important terms rather than seven terms in the
    usual effective Lagrangian of hidden symmetry in
    \cite{M.Bando}\cite{M.Bando1}, three additional new terms (i.e., $\alpha, \beta$ and $\gamma$)
    are necessary in our present more general construction on
    effective Lagrangian via the hidden local symmetry approach.
    Of particular, these three terms are found to nonzero when
    comparing with the effective chiral Lagrangian \cite{libian};
    (ii) Even for the usual six terms with coupling constants
    $\alpha_i, i=1,\cdots,6$, three of the them, $\alpha_4,
    \alpha_5$ and $\alpha_6$, turn out to have different behavior
    when comparing their values yielded from the phenomenological
    analysis in the literature \cite{M.Bando}\cite{M.Bando1}
    with the ones determined from the effective chiral theory\cite{libian}.

   (3).~~The relation $-\alpha_4=\alpha_5=\alpha_6=1$ has
    been taken in the literature \cite{M.Bando}\cite{M.Bando1}
    to accommodate the $\rho-$dominance for
    $a_1\rightarrow\gamma\pi$ decay and cancel the strong
    momentum dependence of the coupling $f_{\rho\pi\pi}$ in the
    absence of $a_1-$meson. While in the effective chiral theory,
    it is seen that $\alpha_4$ is positive with the value $\alpha_4=N_c/(2(\pi
    g)^2)$ and $\alpha_5=\alpha_6=0$.

    (4).~~It is natural to ask why the values $\alpha_4, \alpha_5$ and
    $\alpha_6$ extracted from the two cases are so different, and
    how the cancellation of strong momentum dependence of the
    coupling $f_{\rho\pi\pi}$ and $\rho-$dominance in $a_1\rightarrow\gamma\pi$
    decay can be accommodated in the case with positive value of
    $\alpha_4$ and zero values of $\alpha_5$ and $\alpha_6$. The
    answer is attributed to three additional new terms in our
    present more general construction from hidden local symmetry
    approach. They are found to be nonzero from the effective
    chiral theory and their values are determined from the
    effective chiral theory to be $\alpha=-\gamma=2\beta=\alpha_4=N_c/(2(\pi
    g)^2)$. With these values, it can be shown that the strong
    momentum dependence of $f_{\rho\pi\pi}$ will be cancelled when
    $m^2_\rho=6m^2$ and $g=1/\pi$ due to the existence of
    additional new terms, and the $\rho-$dominance for
    $a_1\rightarrow\gamma\pi$ decay can also be realized
    \cite{libian}.

     (5).~~In comparison with the chiral perturbation theory
    (ChPT)\cite{Gasser}, the new terms $\alpha, \beta$ and
    $\gamma$ are related to the terms $L_1, L_2$ and $L_3$ in
    ChPT. Noticing the algebraic relation
    \begin{eqnarray}
    &&Tr(D_\mu UD_\nu U^\dag D^\mu UD^\nu U^\dag)={1\over2}[Tr(D_\mu UD^\mu
    U^\dag)]^2\nonumber\\
    &&+Tr(D_\mu UD_\nu U^\dag)\cdot Tr(D^\mu UD^\nu U^\dag)-2Tr(D_\mu UD^\mu
    U^\dag)^2\label{algebra1}
    \end{eqnarray}
    we get the relation $L_1=1/2L_2$. According to
    $(\ref{algebra1})$ and the following ($\ref{algebra2}$), we
    can express $L_1, L_2$ and $L_3$ in terms of $\alpha, \beta$ and
    $\gamma$ as:
    \begin{eqnarray}
    L_1={\beta\over2}{1\over12g^2_G},~~~L_2=\beta{1\over12g^2_G},~~~L_3=(\alpha-2\beta+\gamma){1\over12g^2_G},~~~L_9=(3\alpha_4-\alpha){1\over12g^2_G}
    \end{eqnarray}

    (6).~~The terms $\alpha_4$ and $\alpha$ are related to
    the coupling constants $L_9$ in ChPT. Both the sign and
    extracted value for $\alpha_4$ in our present considerations
    are consistent with the ones of $L_9$ from the phenomenology
    well described by ChPT, while the previous results for
    $\alpha_4$ given in literature \cite{M.Bando}\cite{M.Bando1}
    seem to be conflict with the extracted value of $L_9$ in ChPT.

It is then not difficult to show that the more general effective
Lagrangian constructed via the approach of global chiral symmetry
and hidden local chiral symmetry with an appropriate gauge choice
should be consistent with any other effective chiral Lagrangian in
the chiral limit. The fourteen parameters in the effective
Lagrangian up to $O(p^4)$ of the mesons fields can be extracted
from the effective chiral theory.

%%%%%%%%%%%%%%%%%%%%%%%%%%%%%%%%%%%%%%%%%%%%%%%%%%%%%%
\section{ EFFECTIVE CHIRAL LAGRANGIAN AND LOW
ENERGY BEHAVIOR}

A consistent effective Lagrangian should reproduce the low energy
phenomenologies which have been tested by experiments. Now, let's
check the Vector-Pseudoscalar-Pseudoscalar vertex. As an example,
we may first work out the $\rho\pi\pi$ coupling $f_{\rho\pi\pi}$
which is defined as
\begin{eqnarray}
{\cal
L}_{\rho\pi\pi}=f_{\rho\pi\pi}\epsilon_{ijk}\rho_i^\mu\pi_j\partial_\mu\pi_k
\end{eqnarray}

From the general Lagrangian ($\ref{p2}$) and ($\ref{p4}$), it is
easy to get
\begin{eqnarray}
f_{\rho\pi\pi}&=&g_G\bigg\{1+\frac{2m_\rho^2}{g_G^2f^2_\pi}\bigg[(\alpha_4-\alpha/3)\bigg(1-\frac{2c}{a+b+2c}\bigg)^2-\bigg(\frac{2c}{a+b+2c}\bigg)^2\nonumber\\
&&+(\alpha_5+\alpha_6)\bigg(\frac{2c}{a+b+2c}\bigg)\bigg(1-\frac{2c}{a+b+2c}\bigg)\bigg]\bigg\}\label{frpp}
\end{eqnarray}
from the above expression we can see that there is contribution
from $\alpha$. It is seen that when the parameters take the values
chosen from the phenomenological analysis in
\cite{M.Bando}\cite{M.Bando1}, i.e., $a=b=c=2,
-\alpha_4=\alpha_5=\alpha_6=1$ and $\alpha=0$, one has
\begin{eqnarray}
f_{\rho\pi\pi}=g_G=2/g
\end{eqnarray}
where the second term in the curled bracket of ($\ref{frpp}$)
vanishes due to cancellations from various contributions.
Alternatively, we can take other choices, such as the values in
\cite{libian}. As a consequence, we have
\begin{eqnarray}
f_{\rho\pi\pi}={2\over
g}\bigg\{1+\frac{m^2_\rho}{2\pi^2f^2_\pi}\bigg[{N_c\over3}\bigg(1-\frac{6m^2}{m^2_\rho+6m^2}\bigg)^2-\pi^2g^2\bigg(\frac{6m^2}{m^2_\rho+6m^2}\bigg)^2\bigg]\bigg\}
\end{eqnarray}

It is seen that only for a specific choice $g=1/\pi$,
$m^2_\rho=6m^2$ and $N_c=3$, we get $f_{\rho\pi\pi}=2/g$. It can
be shown that with the parameters fixed from the effective chiral
theory, the effective chiral Lagrangian can also lead to a
consistent prediction on $\Gamma(a_1\rightarrow\rho\pi)$ and
$\Gamma(a_1\rightarrow\gamma\pi)$. The numerical result were found
to be $\Gamma(a_1\rightarrow\rho\pi)\simeq 326 MeV$ and
$\Gamma(a_1\rightarrow\gamma\pi)\simeq 252KeV$. In general, the
value of the basic parameter $g$ closing to $1/\pi$ is found to be
a consistent one. Here the term $\alpha$ plays an important role.

The second effect of the additional term $\alpha$ in the more
general effective chiral Lagrangian is that the weinberg's sum
rule $g_a^2=g_\rho^2$ will be modified to be
\begin{eqnarray}
g_a^2=g^2_\rho(1-{\alpha\over3})=g^2_\rho(1-\frac{N_c}{6\pi^2g^2})
\end{eqnarray}
where $g_a$ and $g_\rho$ were defined in \cite{libian}. In the
second equation, the parameter $\alpha$ has been taken the result
fixed from the effective chiral theory. This modification makes
the predictions for the masses of the axial-vectors to be more
consistent with the experimental data.

The third important effect from the additional term $\alpha$ is
the evaluation for the decay constants of the pseudoscalars.

To be more explicit, we may use some algebraic relations and
equation of motion
\begin{eqnarray}
D^\mu(U^\dag D_\mu U)={1\over2}(U^\dag\chi-\chi^\dag
U)-{1\over6}Tr(U^\dag\chi-\chi^\dag U)
\end{eqnarray}
to reexpress the $\alpha$ term into several more familiar terms,
so that its effect can be easily seen. It is easy to check that
\begin{eqnarray}
D_\mu D_\nu UD^\mu D^\nu
U^\dag&=&{1\over2}[F_L^2+F_R^2-2F_LUF_RU^\dag]\nonumber\\
&&+i[D_\mu UD_\nu U^\dag F_L^{\mu\nu}+D_\mu U^\dag D_\nu
UF^{\mu\nu}_R]\nonumber\\
&&+(D_\mu UD^\mu U^\dag)(D_\nu UD^\nu U^\dag)\nonumber\\
&&+{1\over2}D_\mu UD^\mu[U^\dag(U\chi^\dag-\chi^\dag
U^\dag)]\nonumber\\
&&+{1\over2}(D_\mu U^\dag D^\mu U)[U^\dag\chi-\chi^\dag U)]\nonumber\\
&&+\mbox{total derivative terms or trace terms}\label{algebra2}
\end{eqnarray}
From the explicit form, it is not difficult to understand its
effects. Where the first term modifies the Weinberg's sum rule,
the second term contributes to the $\rho\pi\pi$ coupling and the
coupling constant $L_9$ in ChPT, the third term has effects on the
coupling constant $L_3$ in ChPT and the last two terms will
provide additional contributions to the decay constants of
psuedoscalars. As a consequence, we arrive at a complete
prediction for the coupling $L_1, L_2, L_3$ and $L_9$ at this
order, which is consistent with the ones extracted from
phenomenology described by the chiral perturbation theory up to
$O(p^4)$. The numerical values are found to be\\
\begin{center}
\begin{tabular}{|c|c|c|c|c|}
  \hline
  Parameters & $10^3L_1$ & $10^3L_2$ & $10^3L_3$ & $10^3L_9$ \\
  \hline
  Present & $0.79$ & $1.58$ & $-3.16$ & $6.32$ \\
  \hline
  ChPT\cite{Ecker} & $0.4\pm0.3$ & $1.35\pm0.3$ & $-3.5\pm1.1$ & $6.9\pm0.7$ \\
  \hline
\end{tabular}
\end{center}

Now, let's check the known KSFR relation. From the general
effective Lagrangian, the mass of $\rho$ meson is expressed as
$m^2_\rho=af^2_\pi g_G^2/4$. Comparing with the effective chiral
theory with $g^2_G=4/g^2$ and $f_{\rho\pi\pi}\simeq2/g$, one has
\begin{eqnarray}
m^2_\rho=af^2_\pi g^2_G/4={a\over4}f_\pi^2({2\over g})^2\simeq
{a\over4}f^2_\pi f^2_{\rho\pi\pi}
\end{eqnarray}
Thus the known KSFR relation holds for $a\simeq 2$ which is also
consistent with vector meson dominance.

It is seen that the more general effective Lagrangian with its
parameters extracted from the effective chiral theory can well
reproduce the phenomenologies of $\rho-\pi$ system.

The fourth effect of the new terms is the important contributions
to the $\pi-\pi$ scattering
\cite{Gasser}\cite{pion-pion}\cite{libian}.

One may see that only from the $\rho\pi\pi$ coupling,
$a_1\rightarrow\rho\pi$ and $a_1\rightarrow\gamma\pi$ decays, the
parameters appearing in the $O(p^4)$ in the effective chiral
Lagrangian constructed from the hidden local symmetry approach may
not uniquely be determined. The value of parameter $\alpha_4$
extracted from the phenomenology of $\rho-a_1$ system in the
literatures \cite{M.Bando}\cite{M.Bando1} is conflict with the one
from the phenomenology well described by the chiral perturbation
theory and effective chiral theory. While the resulting structure
and couplings from the effective chiral theory are consistent not
only with the phenomenology of $\rho-a_1$ system, but also with
the chiral perturbation theory. Thus, the effective chiral theory
derived from the chiral quarks and bound state solutions of
nonperturbative QCD may provide a very useful way to extract all
the parameters in terms of only two basic scales $m$ and
$f_\pi=186MeV$ (or coupling constant g). It is likely that the
structure of the effective chiral Lagrangians for the $O(p^4)$
given in literatures \cite{M.Bando}\cite{M.Bando1} is incomplete.
As a consequence, the extracted coupling constant are not
reliable.

%%%%%%%%%%%%%%%%%%%%%%%%%%%%%%%%%%%%%%%%%%%%%%%%%%%%%%
\section{CONCLUSIONS}

The more general effective chiral Lagrangian of mesons
(pseudoscalars, vectors and axial-vectors) has been constructed in
the chiral limit by using explicit global chiral symmetry
$U(3)_L\times U(3)_R$ and hidden local chiral symmetry
$\hat{U}(3)_L\times \hat{U}(3)_R$. It is shown that there are many
extra terms in addition to the eleven terms given in paper
\cite{M.Bando}. Among these extra terms there are three important
terms that have been found to play important roles in
understanding the vector meson dominance and the $\pi-\pi$
scattering, in providing consistent predictions on the decay rates
of $a_1\rightarrow\rho\pi$ and $a_1\rightarrow\gamma\pi$, as well
as in resulting a consistent effective chiral Lagrangian with the
chiral perturbation theory.

It is observed that not only the three new interactional terms
introduced in this paper are necessary, but also the resulting
coupling constants for other three interacting terms have total
different values in comparison with the ones given in the
literature from hidden symmetry approach to $O(p^2)$. It is likely
that the structure of the effective Lagrangian to $O(p^4)$ given
in literature \cite{M.Bando} is incomplete, thus the extracted
coupling constants are not reliable.
%%%%%%%%%%%%%%%%%%%%%%%%%%%%%%%%%%%%%%%%%%%%%%%
\acknowledgments

  We would like to thank B.A.Li for his valuable discussions.
This work was supported in part by the key projects of Chinese
Academy of Sciences, the National Science Foundation of China
(NSFC).

%%%%%%%%%%%%%%%%%%%%%%%%%%%%%%%%%%%%%%%%
\appendix

\section{ THE FULL LAGRANGIAN TO $O(p^4)$}

 In general, the $O(p^4)$ lagrangian has two forms which corresponding to one trace operator terms and two trace operator terms.
 One trace operator terms are constructed by basic blocks while two trace operator terms are
 constructed by two $O(p^2)$ terms. As the two trace terms corresponding higher order contributions,
 we will not consider those terms. For convenience, we construct the $O(p^4)$ Lagrangian from the parity properties of the independent fields.\\

(i), terms independent of $\hat{a}_-,
(\hat{a}_{-\mu}\equiv\hat{\xi}_L\hat{a}_{L\mu}\hat{\xi}_L^\dag)$
\begin{eqnarray}
{\cal L}^4_a&=&a_1Tr[(a_{\mu-}a^\mu_-)^2]+a_2Tr[a_{\mu-}a_{\nu-}
a^\mu_- a^\nu_-]+a_3Tr[(a_{\mu+}a^\mu_+)^2]+a_4Tr[a_{\mu+}a_{\nu+}
a^\mu_+ a^\nu_+]\nonumber\\
&&+a_5Tr[a_{\mu-}a^\mu_- a_{\nu+} a^\nu_+]+a_6Tr[a_{\mu-}a_{\nu-}
a^\mu_+ a^\nu_+]+a_7Tr[a_{\mu-}a_{\nu-} a^\nu_+
a^\mu_+]\nonumber\\
&&+a_8\{Tr[a_{\mu-}a^\mu_+ a_{\nu-} a^\nu_+]+Tr[a_{\mu+}a^\mu_-
a_{\nu+} a^\nu_-]\}+a_9Tr[a_{\mu-}a_{\nu+} a^\mu_- a^\nu_+]
\end{eqnarray}

\indent (ii), terms depend on $\hat{a}_-$
\begin{eqnarray}
{\cal L}^4_{\hat{a}}&=&\hat{a}_1Tr[a_{\mu-}a_{\nu-} a^\mu_-
\hat{a}_-^\nu]+\hat{a}_2Tr[a_{\mu-} a^\mu_-
a_{\nu-}\hat{a}_-^\nu]+\hat{a}_3Tr[a_{\nu-}a_{\mu-}a^\mu_- \hat{a}_-^\nu]\nonumber\\
&&+\hat{a}_4Tr[a_{\mu-}a_{\nu-}
\hat{a}_-^\mu\hat{a}_-^\nu]+\hat{a}_5Tr[a_{\mu-}a^\mu_- \hat{a}_{-\nu}\hat{a}_-^\nu]+\hat{a}_6Tr[a_{\mu-}a_{\nu-}\hat{a}_-^\nu\hat{a}_-^\mu]\nonumber\\
&&+\hat{a}_7Tr[a_{\mu-} \hat{a}_-^\mu
a_{\nu-}\hat{a}_-^\nu]+\hat{a}_8Tr[a_{\mu-}
\hat{a}_{-\nu} a^\mu_- \hat{a}_-^\nu]+\hat{a}_9Tr[a_{\mu-}\hat{a}_-^\nu a_{\nu-}\hat{a}_-^\mu]\nonumber\\
&&+\hat{a}_{10}Tr[a_{\mu-}\hat{a}_{-\nu} \hat{a}_-^\mu
\hat{a}_-^\nu]+\hat{a}_{11}Tr[a_{\mu-}\hat{a}_-^\mu
\hat{a}_{-\nu}\hat{a}_-^\nu]+\hat{a}_{12}Tr[a_{\mu-}\hat{a}_{-\nu}\hat{a}_-^\nu\hat{a}_-^\mu]\nonumber\\
&&+\hat{a}_{13}Tr[a_{\mu+}a_{\nu+}
\hat{a}_-^\mu\hat{a}_-^\nu]+\hat{a}_{14}Tr[a_{\mu+}a^\mu_+
\hat{a}_{-\nu}\hat{a}_-^\nu]+\hat{a}_{15}Tr[a_{\mu+}a_{\nu+} \hat{a}_-^\nu\hat{a}_-^\mu]\nonumber\\
&&+\hat{a}_{16}Tr[a_{\mu+}\hat{a}_{-\nu} a^\mu_+
\hat{a}_-^\nu]+\hat{a}_{17}Tr[a_{\mu+}\hat{a}_-^\mu a_{\nu+}
\hat{a}_-^\nu]+\hat{a}_{18}Tr[a_{\mu+}\hat{a}_{-\nu} a^\nu_+
h^\mu]\nonumber\\
&&+\hat{a}_{19}Tr[\hat{a}_{-\mu} a^\mu_- a_{\nu+}
a^\nu_+]+\hat{a}_{20}Tr[\hat{a}_{-\mu} a_{\nu-} a^\mu_+
a^\nu_+]+\hat{a}_{21}Tr[\hat{a}_{-\mu} a^\nu_- a_{\nu+}
a^\mu_+]\nonumber\\
&&+\hat{a}_{22}Tr[a_{\mu-} h^\mu a_{\nu+}
a^\nu_+]+\hat{a}_{23}Tr[a_{\mu-} \hat{a}_{-\nu} a^\nu_+
a^\mu_+]+\hat{a}_{24}Tr[a_{\mu-} \hat{a}_{-\nu} a^\mu_+
a^\nu_+]\nonumber\\
&&+\hat{a}_{25}Tr[a_{\mu-}a_{\nu+} \hat{a}_-^\mu
a^\nu_+]+\hat{a}_{26}Tr[a_{\mu-}a^\nu_+ \hat{a}_{-\nu}
a^\mu_+]+\hat{a}_{27}Tr[a_{\mu-}a^\mu_+ \hat{a}_{-\nu} a^\nu_+]\nonumber\\
&&+\hat{a}_{28}Tr[\hat{a}_{-\mu} \hat{a}_{-\nu} \hat{a}_-^\mu
\hat{a}_-^\nu]+\hat{a}_{29}Tr[(\hat{a}_{-\mu} \hat{a}_-^\mu)^2]
\end{eqnarray}

\indent (iii), terms depend on $V_{\mu\nu}$ and $A_{\mu\nu}$
\begin{eqnarray}
{\cal
L}^4_F&=&\alpha_1(-i/g_G^2)Tr[a_{L\mu}a_{L\nu}F^{L\mu\nu}+a_{R\mu}a_{R\nu}F^{R\mu\nu}]\nonumber\\
&&+\alpha_2(-i/g_G^2)Tr[Ua_{R\mu}a_{R\nu}U^\dag F^{L\mu\nu}+a_{L\mu}a_{L\nu}UF^{R\mu\nu}U^\dag]\nonumber\\
&&+\alpha_3(i/2g_G^2)Tr[a_{L\mu}Ua_{R\nu}U^\dag F^{L\mu\nu}+a_{R\mu}U^\dag a_{L\nu}UF^{R\mu\nu}]+H.c.\nonumber\\
&&+\alpha_4(-i/4g_G^2)Tr[\hat{\xi}_L\hat{a}_{L\mu}\hat{a}_{L\nu}\hat{\xi}_L^\dag F^{L\mu\nu}+\hat{\xi}_R\xi_M^\dag\hat{a}_{L\mu}\hat{a}_{L\nu}\xi_M\hat{\xi}_R^\dag F^{R\mu\nu}]\nonumber\\
&&+\alpha_5(+i/4g_G^2)Tr[a_{L\mu}\hat{\xi}_L\hat{a}_{L\nu}\hat{\xi}_L^\dag F^{L\mu\nu}+a_{R\mu}\hat{\xi}_R\hat{a}_{R\nu}\hat{\xi}_R^\dag F^{R\mu\nu}]+H.c.\nonumber\\
&&+\alpha_6(-i/4g_G^2)Tr[Ua_{R\mu}U^\dag\hat{\xi}_L\hat{a}_{L\nu}\hat{\xi}_L^\dag F^{L\mu\nu}+U^\dag a_{L\mu}U\hat{\xi}_R\hat{a}_{R\nu}\hat{\xi}_R^\dag F^{R\mu\nu}]+H.c.\nonumber\\
\end{eqnarray}

After taking the unitary gauge used in the context, the final
Lagrangian can be easily written down, we shall not list the full
results here.

%%%%%%%%%%%%%%%%%%%%%%%%%%%%%%%%%%%%%%%%%%%%%%%%

\end{document}